# Eliciting Single-Peaked Preferences
# Using Comparison Queries

**Vincent Conitzer**                                               CONITZER@CS.DUKE.EDU
*Departments of Computer Science
and Economics, Duke University
Durham, NC, USA*

## Abstract

Voting is a general method for aggregating the preferences of multiple agents. Each agent ranks all the possible alternatives, and based on this, an aggregate ranking of the alternatives (or at least a winning alternative) is produced. However, when there are many alternatives, it is impractical to simply ask agents to report their complete preferences. Rather, the agents' preferences, or at least the relevant parts thereof, need to be *elicited*. This is done by asking the agents a (hopefully small) number of simple queries about their preferences, such as *comparison* queries, which ask an agent to compare two of the alternatives. Prior work on preference elicitation in voting has focused on the case of unrestricted preferences. It has been shown that in this setting, it is sometimes necessary to ask each agent (almost) as many queries as would be required to determine an arbitrary ranking of the alternatives. In contrast, in this paper, we focus on *single-peaked* preferences. We show that such preferences can be elicited using only a linear number of comparison queries, if either the order with respect to which preferences are single-peaked is known, or at least one other agent's complete preferences are known. We show that using a sublinear number of queries does not suffice. We also consider the case of *cardinally* single-peaked preferences. For this case, we show that if the alternatives' cardinal positions are known, then an agent's preferences can be elicited using only a logarithmic number of queries; however, we also show that if the cardinal positions are not known, then a sublinear number of queries does not suffice. We present experimental results for all elicitation algorithms. We also consider the problem of only eliciting enough information to determine the aggregate ranking, and show that even for this more modest objective, a sublinear number of queries per agent does not suffice for known ordinal or unknown cardinal positions. Finally, we discuss whether and how these techniques can be applied when preferences are *almost* single-peaked.

## 1. Introduction

In multiagent systems, a group of agents often has to make joint decisions even when the agents have conflicting preferences over the alternatives. For example, agents may have different preferences over possible joint plans for the group, allocations of tasks or resources among members of the group, potential representatives (*e.g.*, presidential candidates), *etc.* In such settings, it is important to be able to *aggregate* the agents' individual preferences. The result of this aggregation can be a single alternative, corresponding to the group's collective decision, or a complete aggregate (compromise) ranking of all the alternatives (which can be useful, for instance, if some of the alternatives later turn out not to be feasible). The most general framework for aggregating the agents' preferences is to have the agents *vote* over the alternatives. That is, each agent announces a complete ranking of all alternatives (the agent's *vote*), and based on these votes an outcome (*i.e.*, a winning alternative or a complete aggregate ranking of all alternatives) is chosen according to some *voting rule*.[1]

One might try to create an aggregate ranking as follows: for given alternatives $a$ and $b$, if more votes prefer $a$ to $b$ than vice versa (*i.e.*, $a$ wins its *pairwise election* against $b$), then $a$ should be ranked above $b$ in the aggregate ranking. Unfortunately, when the preferences of the agents are unrestricted and there are at least three alternatives, *Condorcet cycles* may occur. A Condorcet cycle is a sequence of alternatives $a_1, a_2, \ldots, a_k$ such that for each $1 \leq i < k$, more agents prefer $a_i$ to $a_{i+1}$ than vice versa, and more agents prefer $a_k$ to $a_1$ than vice versa. In the presence of a Condorcet cycle, it is impossible to produce an aggregate ranking that is consistent with the outcomes of all pairwise elections. Closely related to this phenomenon are numerous impossibility results that show that every voting rule has significant drawbacks in this general setting. For example, when there are at least three alternatives, Arrow's impossibility theorem (Arrow, 1963) shows that any voting rule for which the relative order of two alternatives in the aggregate ranking is independent of how agents rank alternatives other than these two (*i.e.*, any rule that satisfies *independence of irrelevant alternatives*) must either be *dictatorial* (*i.e.*, the rule simply copies the ranking of a fixed agent, ignoring all other agents) or conflicting with *unanimity* (*i.e.*, for some alternatives $a$ and $b$, the rule sometimes ranks $a$ above $b$ even if all agents prefer $b$ to $a$). As another example, when there are at least three alternatives, the Gibbard-Satterthwaite theorem (Gibbard, 1973; Satterthwaite, 1975) shows that for any voting rule that is *onto* (for every alternative, there exist votes that would make that alternative win) and nondictatorial, there are instances where an agent is best off casting a vote that does not correspond to the agent's true preferences (*i.e.*, the rule is not *strategy-proof*).

## 1.1 Single-Peaked Preferences

Fortunately, these difficulties can disappear if the agents' preferences are restricted, *i.e.*, they display some structure. The best-known, and arguably most important such restriction is that of *single-peaked preferences* (Black, 1948). Suppose that the alternatives are ordered on a line, from left to right, representing the alternatives' *positions*. For example, in a political election, a candidate's position on the line may indicate whether she is a left-wing or a right-wing candidate (and how strongly so). As another example, the alternatives may be numerical values: for example, agents may vote over the size of a budget. As yet another example, the alternatives may be locations along a road (for example, if agents are voting over where to construct a building, or where to meet for dinner, *etc.*). We say that an agent's preferences are *single-peaked* with respect to the alternatives' positions if, on each side of the agent's most preferred alternative (the agent's *peak*), the agent prefers alternatives that are closer to its peak. For example, if the set of alternatives is $\{a, b, c, d, e, f\}$, their positions may be represented by $d < b < e < f < a < c$, in which case the vote $f \succ e \succ b \succ a \succ c \succ d$ is single-peaked, but the vote $f \succ e \succ a \succ d \succ c \succ b$ is not ($b$ and $d$ are on the same side of $f$ in the positions, and $b$ is closer to $f$, so $d$ should not be ranked higher than $b$ if $f$ is the peak). (Throughout, we assume that all preferences are strict, that is, agents are never indifferent between two alternatives.) Preferences are likely to be single-peaked if the alternatives' positions are of primary importance in determining an agent's preferences. For example, in political elections, if voters' preferences are determined primarily by candidates' proximity to their own stance on the left-to-right spectrum, preferences are likely to be single-peaked. If other factors are also important, such as the perceived amicability of the candidates, then preferences are not necessarily likely to be single-peaked.

---

1. One may argue that this approach is not fully general because it does not allow agents to specify their preferences over *probability distributions* over alternatives. For example, it is impossible to know from an agent's vote whether that agent prefers its second-ranked alternative to a 1/2 - 1/2 probability distribution over its first-ranked and third-ranked alternatives. In principle, this can be addressed by voting over these probability distributions instead, although in practice this is usually not tractable.

Formally, if an agent with single-peaked preferences prefers $a_1$ to $a_2$, one of the following must be true:

- $a_1$ is the agent's peak,
- $a_1$ and $a_2$ are on opposite sides of the agent's peak, or
- $a_1$ is closer to the peak than $a_2$.

When all agents' preferences are single-peaked (with respect to the same positions for the alternatives), it is known that there can be no Condorcet cycles. If, in addition, we assume that the number of agents is odd, then no pairwise election can result in a tie. Hence, our aggregate ranking can simply correspond to the outcomes of the pairwise elections. In this case, there is also no incentive for an agent to misreport its preferences, since by reporting its preferences truthfully, it will, in each pairwise election, rank the alternative that it prefers higher.

## 1.2 How Important Are Single-Peaked Preferences?

Before we start developing techniques that deal specifically with single-peaked preferences, we should consider whether this restricted class of preferences is of any interest. The concept of single-peaked preferences has been extremely influential in political science: it is presumably the best-known restriction of preferences there, and it lies at the basis of much of the analytical work in political science. A thorough discussion is given in the book "Analytical Politics" by Hinich and Munger (1997). This book almost immediately jumps into single-peaked preferences, and argues that, in some form, these models date back to Aristotle.

In spite of the importance of single-peaked preferences in the (analytical) political science literature, we are not aware of any empirical studies of how often voters' preferences are actually single-peaked. Some reflection reveals that the answer is probably highly dependent on the particular election. For example, in an election with a clear left-wing candidate, a clear centrist candidate, and a clear right-wing candidate, it seems likely that voters' preferences will in fact be single-peaked: in this case, single-peakedness is equivalent to not ranking the centrist candidate last. However, it is easy to imagine other scenarios where preferences are not single-peaked. For example, the candidates may lie not on a one-dimensional spectrum, but in a two-dimensional space—for instance, a candidate may be a left-wing candidate in terms of social issues but a right-wing candidate in terms of economic issues. Unfortunately, it turns out that multidimensional analogues of single-peaked preferences no longer have the nice properties listed above: Condorcet cycles can occur once again, and strategic misreporting of preferences again becomes an issue.

It seems that in a political context, there will be some settings where preferences are in fact single-peaked, some settings in which they are very close to single-peaked (for example, there may be some voters that take the physical attractiveness of the candidates into account and as a result have preferences that are not quite single-peaked), and some settings in which they are really not single-peaked (for example, because it is a truly multidimensional setting with multiple unrelated issues). We will discuss the generalization of our techniques to "almost" single-peaked preferences in Section 6; multidimensional settings are left for future research.

However, political elections are by no means the only setting of interest, especially from an AI perspective; and in fact, some other important settings seem to fit the single-peaked preference model much better. For example, let us again consider the case where the alternatives are numerical values, such as potential sizes of a budget on which the agents are deciding. Here, it seems quite reasonable that every agent has a most preferred budget size, and always prefers sizes closer to her ideal one. In fact, single-peaked preferences seem more reasonable here than in a political election, where the left-to-right spectrum is merely an approximation of more complex phenomena; in the case of numerical alternatives, however, the order on the alternatives is absolute, and this is likely to be reflected in the preferences.

Even in settings where preferences are not necessarily single-peaked, we could simply *require* the agents to report single-peaked preferences. The merits of this are somewhat debatable, because we are now forcing non-single-peaked agents to misreport their preferences. On the other hand, this does have the benefit of avoiding Condorcet cycles (at least in the reported preferences) so that there is a natural aggregate ranking.[2] A related approach that is sometimes suggested is to frame the issue on which the agents are voting (that is, choose the set of possible alternatives) in such a way as to make single-peaked preferences likely; again, the merits of this are debatable.

## 1.3 Preference Elicitation

A key difficulty in aggregating the preferences of multiple agents is the *elicitation* of the agents' preferences. In many settings, particularly those with large sets of alternatives, having each agent communicate all of its preferences is impractical. For one, it can take up a large amount of communication bandwidth. Perhaps more importantly, in order for an agent to communicate all of its preferences, it must first *determine* exactly what those preferences are. This can be a complex task, especially when no guidance is provided to the agent as to what the key questions are that it needs to answer to determine its preferences.

An alternative approach is for an *elicitor* to sequentially ask the agents certain natural *queries* about their preferences. For example, the elicitor can ask an agent which of two alternatives it prefers (a *comparison query*). Three natural goals for the elicitor are to (1) learn enough about the agents' preferences to determine the winning alternative, (2) learn enough to determine the entire aggregate ranking, and (3) learn each agent's complete preferences. (1) and (2) have the advantage that in general, not all of each agent's preferences need to be determined. For example, for (1), the elicitor does not need to elicit an agent's preferences among alternatives for which we have already determined (from the other agents' preferences) that they have no chance of winning. But even (3) can have significant benefits over not doing any elicitation at all (*i.e.*, having each agent communicate all of its preferences on its own). First, the elicitor provides the agent with a systematic way of assessing its preferences: all that the agent needs to do is answer simple queries. Second, and perhaps more importantly, once the elicitor has elicited the preferences of some agents, the elicitor will have some understanding of which preferences are more likely to occur (and, perhaps, some understanding of why this is so). The elicitor can then use this understanding to guide the elicitation of the next agent's preferences, and learn these preferences more rapidly.

---

2. Analogously, in *combinatorial auctions* (where agents can bid on bundles of items instead of just on individual items) (Cramton et al., 2006), there are often some bundles on which agents are not allowed to bid, for a variety of reasons. Presumably, this still leads to better results than having a separate auction for each item.

In this paper, we study the elicitation of single-peaked preferences using only comparison queries. We mostly focus on approach (3), *i.e.*, learning each agent's complete preferences—though we will show in Section 5 that, at least in the worst case, we cannot do much better if we pursue approach (2), *i.e.*, learning enough about the preferences to determine the aggregate ranking. In this paper, we do not devote much space to (1), *i.e.*, learning enough about the preferences to determine the winning alternative. It should be noted that (1) is a significantly easier objective: it is well known that the (Condorcet) winner is always the median of the agents' most preferred alternatives, so that it suffices to find each agent's most preferred alternative (see the note at the end of Section 5). For most of the paper, we assume that preferences are always single-peaked (the exception is Section 6 in which we discuss the case where preferences are usually, but not always, single-peaked).

In Section 3, we study single-peaked preferences in their general form as described in Subsection 1.1 (also known as *ordinally* single-peaked preferences, in contrast to the *cardinally* single-peaked preferences studied later in the paper). We study both the setting where the elicitor knows the positions of the alternatives (Subsection 3.1), and the setting where the elicitor (at least initially) does not (Subsection 3.2). Experimental results for this section are provided in Appendix A. Then, in Section 4, we study the more restricted setting where preferences are *cardinally* single-peaked—that is, each alternative and each agent has a cardinal position in $\mathbb{R}$, and agents rank alternatives by distance to their own cardinal position. (To prevent confusion, we emphasize that before Section 4, an alternative's position only refers to its place in the order of alternatives, *not* to a real number.) Again, we study both the setting where the elicitor knows the positions of the alternatives (Subsection 4.1), and the setting where the elicitor (at least initially) does not (Subsection 4.2). Experimental results for this section are provided in Appendix B. All our elicitation algorithms completely elicit one agent's preferences before moving on to the next agent (as opposed to going back and forth between agents). This gives the algorithms a nice online property: if agents arrive over time, then we can elicit an agent's preferences when it arrives, after which the agent is free to leave (as opposed to being forced to wait until the arrival of the next agent). Especially in the case where our goal is just to find the aggregate ranking, one may wonder if it can be more efficient to go back and forth between agents. It turns out that, at least in the worst case, this cannot help (significantly), by a result presented in Section 5.

The following tables summarize the results in this paper (with the exception of the experimental results in Appendices A and B, which illustrate how the algorithms perform on random instances, and the results in Section 6 for preferences that are *almost* single-peaked).

| | positions known | positions unknown |
|---|---|---|
| **ordinal** | $\Theta(m)$ (Subsection 3.1) | $\Theta(m)$ (Subsections 3.1, 3.2) |
| **cardinal** | $\Theta(\log m)$ (Subsection 4.1) | $\Theta(m)$ (Subsections 3.2, 4.2) |

*Table 1: Number of comparison queries required to fully elicit one agent's single-peaked preferences over $m$ alternatives, in the worst case. For the upper bounds in the **positions unknown** column, at least one other agent's preferences must be known (otherwise, there is no restriction on the current agent's preferences and hence the answer is $\Theta(m \log m)$).*

| | positions known | positions unknown |
|---|---|---|
| **ordinal** | $\Theta(nm)$ (Subsection 3.1, Section 5) | $\Theta(nm)$ (Subsection 3.2, Section 5) |
| **cardinal** | $O(n \log m)$ (Subsection 4.1) | $\Theta(nm)$ (Subsection 3.2, Section 5) |

*Table 2: Number of comparison queries required to find the aggregate ranking of $m$ alternatives for $n$ agents with single-peaked preferences, in the worst case. For the upper bounds in the **positions unknown** column, at least one agent's preferences must be known (otherwise, the first agent's preferences can be elicited using $O(m \log m)$ queries).*

## 2. Related Research and the Case of Unrestricted Preferences

In this section, we first discuss related research; then, we make some basic observations about eliciting general (not necessarily single-peaked) preferences, which will serve as a useful baseline for comparison.

### 2.1 Related Research

Voting techniques are drawing increasing interest from the artificial intelligence community, especially from multiagent systems researchers. Voting has been used in applications such as collaborative filtering, for example, by Pennock, Horvitz, and Giles (2000); and planning among multiple automated agents, for example, by Ephrati and Rosenschein (1991, 1993). Some key research topics include voting rules that are computationally hard to execute, for example, by Bartholdi, Tovey, and Trick (1989), Hemaspaandra, Hemaspaandra, and Rothe (1997), Cohen, Schapire, and Singer (1999), Rothe, Spakowski, and Vogel (2003), and Conitzer (2006); voting rules that are computationally hard to *manipulate* by strategic agents, for example, by Bartholdi and Orlin (1991), Hemaspaandra and Hemaspaandra (2007), Procaccia and Rosenschein (2007), and Conitzer, Sandholm, and Lang (2007); and concisely representing preferences in voting (Lang, 2007). Single-peaked preferences have also been studied from a computational-complexity viewpoint (Walsh, 2007).

Preference elicitation is also an important research topic in artificial intelligence; prominent examples of such research include the work of Vu and Haddawy (1997), Chajewska, Getoor, Norman, and Shahar (1998), Vu and Haddawy (1998), Chajewska, Koller, and Parr (2000), Boutilier (2002), and Braziunas and Boutilier (2005), and this list is by no means exhaustive. As for preference elicitation in multiagent systems, a significant body of work focuses on combinatorial auctions—for an overview of this work, see the chapter by Sandholm and Boutilier (2006). Much of this work focuses on elicitation approach (1), *i.e.*, learning enough about the bidders' valuations to determine the optimal allocation. Sometimes, additional information must be elicited from the bidders to determine the payments that they should make according to the Clarke (1971), or more generally, a Groves (1973), mechanism. Example elicitation approaches include ascending combinatorial auctions—for an overview, see the chapter by Parkes (2006)—as well as frameworks in which the auctioneer can ask queries in a more flexible way (Conen & Sandholm, 2001). A significant amount of the research on preference elicitation in combinatorial auctions is also devoted to elicitation approach (3), *i.e.* learning an agent's complete valuation function. In this research, typically valuation functions are assumed to lie in a restricted class, and given this it is shown that an agent's complete valuation function can be elicited using a polynomial number of queries of some kind. Various results of this nature have been obtained by Zinkevich, Blum, and Sandholm (2003), Blum, Jackson, Sandholm, and Zinkevich (2004), Lahaie and Parkes (2004), and Santi, Conitzer, and Sandholm (2004). The work by Lahaie and Parkes (2004) also includes results where the goal is only to determine the outcome.

There has also already been some work on preference elicitation in voting (the setting of this paper). All of the earlier work has focused on elicitation approach (1), eliciting enough information from the agents to determine the winner, without any restriction on the space of possible preferences. Conitzer and Sandholm (2002) studied the complexity of deciding whether enough information has been elicited to declare a winner, as well as the complexity of choosing which votes to elicit given very strong suspicions about how agents will vote. They also studied what additional opportunities for strategic misreporting of preferences elicitation introduces, as well as how to avoid introducing these opportunities. (Strategic misreporting is not a significant concern in the setting of this paper: under the restriction of single-peaked preferences, reporting truthfully is a dominant strategy when agents simultaneously report their complete preferences, and hence responding truthfully to the elicitor's queries is an *ex-post* equilibrium. As such, in this paper we will make no distinction between an agent's vote and its true preferences.) Conitzer and Sandholm (2005) studied elicitation algorithms for determining the winner under various voting rules (without any suspicion about how agents will vote), and gave lower bounds on the worst-case amount of information that agents must communicate. Most recently—after the AAMAS-07 version of this paper (Conitzer, 2007)—the communication complexity of determining the winner or aggregate ranking in domains with single-peaked preferences has been studied (Escoffier et al., 2008). The communication-complexity part of that work (which is not the main contribution of that paper) considers only the ordinal case with known positions, and its results build on the AAMAS-07 version of this paper, as well as on the communication complexity paper mentioned above (Conitzer & Sandholm, 2005).

## 2.2 Eliciting General Preferences

As a basis for comparison, let us first analyze how difficult it is to elicit arbitrary (not necessarily single-peaked) preferences using comparison queries. We recall that our goal is to extract the agent's complete preferences, *i.e.*, we want to know the agent's exact ranking of all $m$ alternatives. This is exactly the same problem as that of sorting a set of $m$ elements, when only binary comparisons between elements can be used to do the sorting. This is an extremely well-studied problem, and it is well-known that it can be solved using $O(m \log m)$ comparisons, for example using the MergeSort algorithm (which splits the set of elements into two halves, solves each half recursively, and then merges the solutions using a linear number of comparisons). It is also well-known that $\Omega(m \log m)$ comparisons are required (in the worst case). One way to see this is that there are $m!$ possible orders, so that an order encodes $\log(m!)$ bits of information—and $\log(m!)$ is $\Omega(m \log m)$. Hence, in general, *any* method for communicating an order (not just methods based on comparison queries) will require $\Omega(m \log m)$ bits (in the worst case).

Interestingly, for some common voting rules (including Borda, Copeland, and Ranked Pairs), it can be shown using techniques from communication complexity theory that even just determining whether a given alternative is the winner requires the communication of $\Omega(nm \log m)$ bits (in the worst case), where $n$ is the number of agents (Conitzer & Sandholm, 2005). That is, even if we do not try to elicit agents' complete preferences, (in the worst case) it is impossible to do more than a constant factor better than having each agent communicate all of its preferences! These lower bounds even hold for *nondeterministic* communication, but they do assume that preferences are unrestricted. In contrast, by assuming that preferences are single-peaked, we can elicit an agent's *complete* preferences using only $O(m)$ queries, as we will show in this paper. Of course, once we know the agents' complete preferences, we can execute any voting rule. This shows how useful it can be for elicitation to know that agents' preferences lie in a restricted class.

## 3. Eliciting Ordinally Single-Peaked Preferences

In this section, we study the elicitation of *ordinally* single-peaked preferences (the general form of single-peaked preferences as described in Subsection 1.1, in contrast to *cardinally* single-peaked preferences, which we study in Section 4). We first study the case where the alternatives' positions are known (Subsection 3.1), and then the case where they are not known (Subsection 3.2). Experimental results for the algorithms in this section are given in Appendix A; these results give some idea of how the algorithms behave on randomly drawn instances, rather than in the worst case.

### 3.1 Eliciting with Knowledge of Alternatives' Ordinal Positions

In this subsection, we focus on the setting where the elicitor knows the positions of the alternatives. Let $p : \{1, \ldots, m\} \rightarrow A$ denote the mapping from positions to alternatives, *i.e.*, $p(1)$ is the leftmost alternative, $p(2)$ is the alternative immediately to the right of $p(1)$, ..., and $p(m)$ is the rightmost alternative. Our algorithms make calls to the function $\mathsf{Query}(a_1, a_2)$, which returns *true* if the agent whose preferences we are currently eliciting prefers $a_1$ to $a_2$, and *false* otherwise. (Since one agent's preferences are elicited at a time, we need not specify which agent is being queried.)

The first algorithm serves to find the agent's peak (most preferred alternative). The basic idea of this algorithm is to do a binary search for the peak. To do so, we need to be able to assess whether the peak is to the left or right of a given alternative $a$. We can discover this by asking whether the alternative immediately to the right of $a$ is preferred to $a$: if it is, then the peak must be to the right of $a$, otherwise, the peak must be to the left of, or equal to, $a$.

---

FindPeakGivenPositions($p$)

$l \leftarrow 1$
$r \leftarrow m$
**while** $l < r$ {
  $m_1 \leftarrow \lfloor (l+r)/2 \rfloor$
  $m_2 \leftarrow m_1 + 1$
  **if** Query($p(m_1), p(m_2)$)
    $r \leftarrow m_1$
  **else**
    $l \leftarrow m_2$
}
**return** $l$

---

Once we have found the peak, we can continue to construct the agent's ranking of the alternatives as follows. We know that the agent's second-ranked alternative must be either the alternative immediately to the left of the peak, or the one immediately to the right. A single query will settle which one is preferred. Without loss of generality, suppose the left alternative was preferred. Then, the third-ranked alternative must be either the alternative immediately to the left of the second-ranked alternative, or the alternative immediately to the right of the peak. Again, a single query will suffice—*etc*. Once we have determined the ranking of either the leftmost or the rightmost alternative, we can construct the remainder of the ranking without asking any more queries (by simply ranking the remaining alternatives according to proximity to the peak). The algorithm is formalized below. It uses the function Append($a_1, a_2$), which makes $a_1$ the alternative that immediately succeeds $a_2$ in the current ranking (*i.e.*, the current agent's preferences as far as we have constructed them). In the pseudocode, we will omit the (simple) details of maintaining such a ranking as a linked list. The algorithm returns the highest-ranked alternative; this is to be interpreted as including the linked-list structure, so that effectively the entire ranking is returned. $c$ is always the alternative that is ranked last among the currently ranked alternatives.

```
FindRankingGivenPositions(p)

t ← FindPeakGivenPositions(p)
s ← p(t)
l ← t − 1
r ← t + 1
c ← s
while l ≥ 1 and r ≤ m {
  if Query(p(l), p(r)) {
    Append(p(l), c)
    c ← p(l)
    l ← l − 1
  } else {
    Append(p(r), c)
    c ← p(r)
    r ← r + 1
  }
}
while l ≥ 1 {
  Append(p(l), c)
  c ← p(l)
  l ← l − 1
}
while r ≤ m {
  Append(p(r), c)
  c ← p(r)
  r ← r + 1
}
return s
```

**Theorem 1** FindRankingGivenPositions *correctly determines an agent's preferences, using at most* $m - 2 + \lceil \log m \rceil$ *comparison queries.*

**Proof**: Correctness follows from the arguments given above. FindPeakGivenPositions requires at most $\lceil \log m \rceil$ comparison queries. Every query after this allows us to add an additional alternative to the ranking, and for the last alternative we will not need a query, hence there can be at most $m - 2$ additional queries. □

Thus, the number of queries that the algorithm requires is linear in the number of alternatives. It is impossible to succeed using a sublinear number of queries, because an agent's single-peaked preferences can encode a linear number of bits, as follows. Suppose the alternatives' positions are as follows: $a_{m-1} < a_{m-3} < a_{m-5} < \ldots < a_4 < a_2 < a_1 < a_3 < a_5 < \ldots < a_{m-4} < a_{m-2} < a_m$. Then, any vote of the form $a_1 \succ \{a_2, a_3\} \succ \{a_4, a_5\} \succ \ldots \succ \{a_{m-1}, a_m\}$ (where the set notation indicates that there is no constraint on the preference between the alternatives in the set, that is, $\{a_i, a_{i+1}\}$ can be replaced either by $a_i \succ a_{i+1}$ or $a_{i+1} \succ a_i$) is single-peaked with respect to the alternatives' positions. The agent's preference between alternatives $a_i$ and $a_{i+1}$ (for even $i$) encodes a single bit, hence the agent's complete preferences encode $(m-1)/2$ bits. Since the answer to a comparison query can communicate only a single bit of information, it follows that a linear number of queries is in fact necessary.

## 3.2 Eliciting without Knowledge of Alternatives' Ordinal Positions

In this subsection, we study a more difficult question: how hard is it to elicit the agents' preferences when the alternatives' positions are not known? Certainly, it would be desirable to have elicitor software that does not require us to enter domain-specific information (namely, the positions of the alternatives) before elicitation begins, for two reasons: (1) this information may not be available to the entity running the election, and (2) entering this information may be perceived by agents as unduly influencing the process, and perhaps the outcome, of the election. Rather, the software should *learn* (relevant) information about the domain from the elicitation process itself.

It is clear that this learning will have to take place over the process of eliciting the preferences of multiple agents. Specifically, without any knowledge of the positions of the alternatives, the first agent's preferences could be *any* ranking of the alternatives, since any ranking is single-peaked with respect to some positions. Hence, eliciting the first agent's preferences will require $\Omega(m \log m)$ queries. Once the elicitor knows the first agent's preferences, though, some ways in which the alternatives may be positioned will be eliminated (but many will remain).

Can the elicitor learn the exact positions of the alternatives? The answer is no, for several reasons. First of all, we can invert the positions of the alternatives, making the leftmost alternative the rightmost, *etc.*, without affecting which preferences are single-peaked with respect to these positions. This is not a fundamental problem because the elicitor could choose either one of the positionings. More significantly, the agents' preferences may simply not give the elicitor enough information to determine the positions. For example, if all agents turn out to have the same preferences, the elicitor will never learn anything about the alternatives' positions beyond what was learned from the first agent. In this case, however, the elicitor could simply try to verify that the next agent whose preferences are to be elicited has the same preferences, which can be done using only a linear number of queries. More generally, one might imagine an intricate elicitation scheme which *either* requires few queries to elicit an agent's preferences, *or* learns something new and useful from these preferences that will shorten the elicitation process for later agents. Then, one might imagine a complex accounting scheme, in the spirit of amortized analysis, showing that the total elicitation cost over many agents cannot be too large.

Fortunately, it turns out that we do not need anything so complex. In fact, knowing even *one* agent's (complete) preferences is enough to elicit any other agent's preferences using only a linear number of queries! (And a sublinear number will not suffice, since we already showed that a linear number is necessary even if we know the alternatives' positions. Hence, no matter how many agents' preferences we already know, we will always require linearly many queries for the next agent.) To prove this, we will give an elicitation algorithm that takes as input one (the first) agent's preferences (*not* the positions of the alternatives), and elicits another agent's preferences using a linear number of queries. Of course, the algorithm can still be applied when we already know the preferences of more than one agent; in that case, we can use the preferences of any one of those agents as input. In the worst case, knowing the preferences of more than one agent will not help us, because it may be that all of those agents have the same preferences, in which case this does not teach us anything about the alternatives' positions beyond what we learned from the first agent. In general (not in the worst case), we may be able to learn something from the additional agents' preferences; however, at best, we learn the alternatives' positions exactly, and even in that case we still need a linear number of queries. Hence, in this paper, we do not investigate how we can use the known preferences of multiple agents.

First, we need a subroutine for finding the agent's peak. We cannot use the algorithm FindPeakGivenPositions from the previous subsection, since we do not know the positions. However, even the trivial algorithm that examines the alternatives one by one and maintains the most preferred alternative so far requires only a linear number of queries, so we will simply use this algorithm.

```
FindPeak()

s ← a₁
for all a ∈ {a₂, …, aₘ}
  if Query(a, s)
    s ← a
return s
```

Once we have found the agent's peak, we next find the alternatives that lie between this peak, and the peak of the known vote (*i.e.*, the peak of the agent whose preferences we know). The following lemma is the key tool for doing so.

**Lemma 1** *Consider votes $v_1$ and $v_2$ with peaks $s_1$ and $s_2$, respectively. Then, an alternative $a \notin \{s_1, s_2\}$ lies between the two peaks if and only if both $a \succ_{v_1} s_2$ and $a \succ_{v_2} s_1$.*

**Proof**: If $a$ lies between the two peaks, then for each $i$, $a$ lies closer to $s_i$ than $s_{3-i}$ (the other vote's peak) lies to $s_i$. Hence $a \succ_{v_1} s_2$ and $a \succ_{v_2} s_1$. Conversely, $a \succ_{v_i} s_{3-i}$ implies that $a$ lies on the same side of $s_{3-i}$ as $s_i$ (otherwise, $v_i$ would have ranked $s_{3-i}$ higher). But since this is true for both $i$, it implies that $a$ must lie between the peaks. □

Thus, to find the alternatives between the peak of the known vote and the peak of the current agent, we simply ask the current agent, for each alternative that the known vote prefers to the peak of the current agent, whether it prefers this alternative to the known vote's peak. If the answer is positive, we add the alternative to the list of alternatives between the peaks.

The two votes must rank the alternatives between their peaks in the exact opposite order. Thus, at this point, we know the current agent's preferences over the alternatives that lie between its peak and the peak of the known vote (including the peaks themselves). The final and most complex step is to integrate the remaining alternatives into this ranking. (Some of these remaining alternatives may be ranked higher than some of the alternatives between the peaks.) The strategy will be to integrate these alternatives into the current ranking one by one, in the order in which the known vote ranks them, starting with the one that the known vote ranks highest. When integrating such an alternative, we first have the current agent compare it to the worst-ranked alternative already in the ranking. We note that the *known* vote must prefer the latter alternative, because this latter alternative is either the known vote's peak, or an alternative that we integrated earlier and that was hence preferred by the known vote. If the latter alternative is also preferred by the current agent, we add the new alternative to the bottom of the current ranking and move on to the next alternative. If not, then we learn something useful about the positions of the alternatives, namely that the new alternative lies on the *other side* of the current agent's peak from the alternative currently ranked last. The following lemma proves this. In it, $v_1$ takes the role of the known vote, $v_2$ takes the role of the agent whose preferences we are currently eliciting, $a_1$ takes the role of the alternative currently ranked last by the current agent, and $a_2$ takes the role of the alternative that we are currently integrating.

**Lemma 2** *Consider votes $v_1$ and $v_2$ with peaks $s_1$ and $s_2$, respectively. Consider two alternatives $a_1, a_2 \neq s_2$ that do* not *lie between $s_1$ and $s_2$. Suppose $a_1 \succ_{v_1} a_2$ and $a_2 \succ_{v_2} a_1$. Then, $a_1$ and $a_2$ must lie on opposite sides of $s_2$.*

**Proof**: If $a_1$ and $a_2$ lie on the same side of $s_2$—without loss of generality, the left side— then, because neither lies between $s_1$ and $s_2$, they must also both lie on the left side of $s_1$ (possibly, one of them is equal to $s_1$). But then, $v_1$ and $v_2$ cannot disagree on which of $a_1$ and $a_2$ is ranked higher. $\square$

For the purpose of integrating this new alternative, knowing that it is on the other side from the alternative currently ranked last is not that helpful. In fact, if we reach this point in the algorithm, we will simply start comparing the new alternative to the alternatives that are already in the ranking, one by one. The benefit of knowing that it is on the other side is that it will make later alternatives easier to integrate: specifically, once we have integrated the new alternative, we know that *all alternatives that we integrate later must end up ranked below this alternative*. This is because of the following lemma, in which $v_1$ takes the role of the known vote, $v_2$ takes the role of the agent whose preferences we are currently eliciting, $a_1$ takes the role of the alternative currently ranked last by the current agent, $a_2$ takes the role of the alternative that we are currently integrating, and $a_3$ takes the role of an alternative to be integrated later.

**Lemma 3** *Consider votes $v_1$ and $v_2$ with peaks $s_1$ and $s_2$, respectively. Consider three alternatives $a_1, a_2, a_3 \neq s_2$ that do* not *lie between $s_1$ and $s_2$. Suppose $a_1 \succ_{v_1} a_2 \succ_{v_1} a_3$, and $a_2 \succ_{v_2} a_1$. Then, $a_2 \succ_{v_2} a_3$.*

**Proof**: By Lemma 2, we know that $a_1$ and $a_2$ must lie on opposite sides of $s_2$. Moreover, because they do not lie between $s_1$ and $s_2$, and are not equal to $s_2$, they must also lie on opposite sides of $s_1$ (with the additional possibility that $a_1 = s_1$). Because $a_1 \succ_{v_1} a_2 \succ_{v_1} a_3$, it follows that either $a_3$ lies on the same side of the peaks as $a_2$, but further away; or on the same side as $a_1$, but further away. In the former case, we immediately obtain $a_2 \succ_{v_2} a_3$; in the latter case, we have $a_1 \succ_{v_2} a_3$, which implies $a_2 \succ_{v_2} a_3$ because $a_2 \succ_{v_2} a_1$.     □

Based on this observation, in the algorithm, we let $c_1$ be the alternative for which we know that all later alternatives that we integrate will end up ranked below it. (In a sense, $c_1$ takes the role of $a_2$ in the above lemmas.) We also keep track of $c_2$, the alternative currently ranked last. (In a sense, $c_2$ takes the role of $a_1$ in the above lemmas.) Then, when we integrate a new alternative, first we compare it to $c_2$; if it is ranked below $c_2$, then we place it below $c_2$, and make it the new $c_2$. If it is ranked above $c_2$, then we compare it to the successor of $c_1$; if it is ranked below that, then we compare it to the successor of the successor; and so on, until we find its place; finally, we make it the new $c_1$. While this last stage may require up to a linear number of queries, the key observation is that *each time we ask such a query, $c_1$ moves down in the ranking by one place*. Hence, the *total* number of such queries that we can ask (summing over *all* alternatives that we integrate) is at most $m-1$, because $c_1$ can move down at most $m-1$ times; and this is why we can get the linear bound on the number of queries.

**Example 1** *Suppose the alternatives' positions are $e < c < b < f < a < d$. Also suppose that the known vote is $a \succ d \succ f \succ b \succ c \succ e$, and the preferences of the agent that we are currently eliciting are $c \succ e \succ b \succ f \succ a \succ d$. The algorithm will proceed as follows. First,* FindPeak *will identify $c$ as the current agent's peak. Now, the alternatives that the known vote prefers to $c$ are $a$ (the known vote's peak), as well as $d, f, b$. For each of the last three alternatives, the algorithm queries the current agent whether it is preferred to $a$. The answer will be positive (only) for $b$ and $f$, so we know that these two alternatives must lie between the peaks $a$ and $c$ on the line (and hence must be ranked oppositely by the known vote and the current agent). At this point, we know that the current agent must prefer $c \succ b \succ f \succ a$, and the algorithm must integrate $d$ and $e$ into this ranking. We set $c_1 = c$ and $c_2 = a$. The algorithm first integrates $d$ since it is ranked higher than $e$ in the known vote. The algorithm queries the agent with $d$ and $c_2 = a$, and $a$ is preferred. Now we know that the current agent must prefer $c \succ b \succ f \succ a \succ d$, and the algorithm sets $c_2 = d$. Finally, the algorithm must integrate $e$. The algorithm queries the agent with $e$ and $c_2 = d$, and $e$ is preferred. The algorithm then queries the agent with $e$ and the successor of $c_1$, which is $b$. $e$ is preferred, so the algorithm inserts $e$ between $c_1 = c$ and $b$, and sets $c_1 = e$. At this point we know the entire ranking $c \succ e \succ b \succ f \succ a \succ d$.*

We now present the algorithm formally. The algorithm again uses the function Append$(a_1, a_2)$, which makes $a_1$ the alternative that immediately succeeds $a_2$ in the current ranking. It also uses the function InsertBetween$(a_1, a_2, a_3)$, which inserts $a_1$ between $a_2$ and $a_3$ in the current ranking. The algorithm will (eventually) set $m(a)$ to *true* if $a$ lies between the peaks of the current agent and the known vote $v$, or if $a$ is the peak of $v$; otherwise, $m(a)$ is set to *false*. $v(i)$ returns the alternative that the known vote ranks $i$th (and hence $v^{-1}(a)$ returns the ranking of alternative $a$ in the known vote, and $v^{-1}(a_1) < v^{-1}(a_2)$ means that $v$ prefers $a_1$ to $a_2$). $n(a)$ returns the alternative immediately following $a$ in the current ranking. Again, only the peak is returned, but this includes the linked-list structure and hence the entire ranking.

```
FindRankingGivenOtherVote(v)

s ← FindPeak()
for all a ∈ A
  m(a) ← false
for all a ∈ A − {s, v(1)}
  if v⁻¹(a) < v⁻¹(s)
    if Query(a, v(1))
      m(a) ← true
c₁ ← s
c₂ ← s
m(v(1)) ← true
for i = m to 1 step −1 {
  if m(v(i)) = true {
    Append(v(i), c₂)
    c₂ ← v(i)
  }
}
for i = 1 to m
  if not (m(v(i)) or v(i) = s)
    if Query(c₂, v(i)) {
      Append(v(i), c₂)
      c₂ ← v(i)
    } else {
      while Query(n(c₁), v(i))
        c₁ ← n(c₁)
      InsertBetween(v(i), c₁, n(c₁))
      c₁ ← v(i)
    }
return s
```

**Theorem 2** FindRankingGivenOtherVote *correctly determines an agent's preferences, using at most $4m - 6$ comparison queries.*

**Proof**: Correctness follows from the arguments given above. FindPeak requires $m - 1$ comparison queries. The next stage, discovering which alternatives lie between the current agent's peak and the known vote's peak, requires at most $m - 2$ queries. Finally, we must count the number of queries in the integration step. This is more complex, because integrating one alternative (which we may have to do up to $m - 2$ times) can require multiple queries. Certainly, the algorithm will ask the agent to compare the alternative currently being integrated to the current $c_2$. This contributes up to $m - 2$ queries in total. However, if the current alternative is preferred over $c_2$, we must ask more queries, comparing the current alternative to the alternative currently ranked immediately behind the current $c_1$ (perhaps multiple times). But every time that we ask such a query, $c_1$ changes to another alternative, and this can happen at most $m-1$ times in total. It follows that the total number of such queries that we ask (summed over *all* the alternatives that we integrate) is at most $m-1$. Adding up all of these bounds, we get a total bound of $(m-1) + (m-2) + (m-2) + (m-1) = 4m-6$. $\quad\square$

Of course, it is impossible to succeed with a sublinear number of queries, for the same reason as in Subsection 3.1. In practice, the algorithm ends up requiring on average roughly $3m$ queries, as can be seen in Appendix A.

## 4. Eliciting Cardinally Single-Peaked Preferences

In this section, we restrict the space of allowed preferences slightly further. We assume that each alternative $a$ has a *cardinal* position $r(a) \in \mathbb{R}$ (as opposed to the merely *ordinal* positions that we have considered up to this point in the paper, which do not provide a distance function over the alternatives). We have already mentioned some examples for which this is reasonable: agents can be voting over a budget, or over a location along a road (in which case an alternative's cardinal position equals its distance from the beginning of the road). Even when the alternatives are political candidates, it may be reasonable for them to have cardinal positions: for example, if the candidates have voting records, then a candidate's cardinal position can be the percentage of times that she cast a "right-wing" vote. Additionally, we assume that every *agent* also has a cardinal position $r$, and that it ranks the alternatives in order of proximity to its own position. That is, the agent sorts the alternatives according to $|r - r(a)|$ (preferring the closer ones). Such preferences are known as *cardinally* single-peaked preferences (again, as opposed to the *ordinally* single-peaked preferences that we have considered up to this point) (Brams et al., 2002, 2005).[3]

Cardinally single-peaked preferences are always ordinally single-peaked as well (with respect to the order $<$ defined by $a < a' \Leftrightarrow r(a) < r(a')$), but the converse does not hold: it is possible for agents' preferences to be ordinally single-peaked, but not cardinally single-peaked. That is, it is possible that all agents' preferences are consistent with a single order of the alternatives, but that there is no way to assign cardinal positions to the alternatives and the agents so that the agents' preferences are given by ranking the alternatives according to proximity to their own cardinal position. The following example shows this:

**Example 2** *Suppose agent 1 prefers $a \succ b \succ c \succ d$, agent 2 prefers $b \succ c \succ d \succ a$, agent 3 prefers $c \succ b \succ a \succ d$. These preferences are all consistent with the order $a < b < c < d$, so they are ordinally single-peaked. Nevertheless, there are no cardinal positions with which they are consistent. We prove this by contradiction: let us suppose there were such cardinal positions. $a$ and $d$ are both sometimes ranked last, which implies they must be the leftmost and rightmost alternatives (without loss of generality, suppose $a$ is the leftmost alternative). Agent 1's preferences then imply that the order must be $a < b < c < d$. For $i \in \{1, 2, 3\}$, let $r(i)$ be agent i's cardinal position. By the agents' preferences, we must have $|r(2) - r(d)| < |r(2) - r(a)|$ and $|r(3) - r(d)| > |r(3) - r(a)|$. Therefore, it must be that $r(2) > r(3)$. On*

---

*the other hand, we must have* $|r(2) - r(b)| < |r(2) - r(c)|$ *and* $|r(3) - r(b)| > |r(3) - r(c)|$. *Therefore, it must be that* $r(2) < r(3)$—*but now we have a contradiction. Hence the above profile of preferences is not cardinally single-peaked.*

A similar example is given by Brams et al. (2002)—in fact, the preferences in the above example are a proper subset of the ones used in their example—although in their proof that the example's preferences are not cardinally single-peaked, they make use of the fact that agents and alternatives coincide in their model, which simplifies the argument somewhat. So, the restriction to cardinally single-peaked preferences does come at a loss in what preferences we can represent. In particular, it should be emphasized that even if alternatives correspond to numerical values, it may very well still be the case that preferences are ordinally, but not cardinally, single-peaked. For example, if agents are voting over a budget for a project, then one agent's preferences might be that the budget should definitely not go above 500 (perhaps because then it starts coming at the cost of another project that the agent cares about); but as long as this constraint is met, then the budget should be as large as possible. Hence, the agent's preferences would be $500 \succ 499 \succ 498 \succ \ldots \succ 1 \succ 0 \succ 501 \succ 502 \succ \ldots$, so they would be ordinally, but not cardinally, single-peaked. Thus, cardinally single-peaked preferences are strictly less likely to occur in practice than ordinally single-peaked preferences; nevertheless, cardinally single-peaked preferences (or something very close to it) may well occur in practice. For example, when the agents are voting over the location of a project along a road, they are likely to rank the locations simply by the distance to their own positions along the road. A more detailed discussion of the symmetry assumption inherent in cardinally single-peaked preferences is given by Hinich and Munger (1997).

Just as we did for ordinally single-peaked preferences, we first study the case where the alternatives' cardinal positions are known (Subsection 4.1), and then the case where they are not known (Subsection 4.2). Experimental results for known cardinal positions are given in Appendix B; these results give some idea of how the algorithm behaves on randomly drawn instances, rather than in the worst case. (We have no experimental results for unknown cardinal positions, because we only have a negative result there.)

## 4.1 Eliciting with Knowledge of Alternatives' Cardinal Positions

In this subsection, we show that if preferences are cardinally single-peaked, and in addition the cardinal positions of the alternatives are known, then there is a preference elicitation algorithm that uses only a *logarithmic* number of comparison queries per agent. At a high level, the algorithm works as follows. For each pair of alternatives $a, a'$ ($r(a) < r(a')$), consider their *midpoint* $m(a, a') = (r(a) + r(a'))/2$. Now, if an agent prefers $a$ to $a'$, that agent's position $r$ must be smaller than $m(a, a')$; otherwise, $r > m(a, a')$. (As before, we assume that there are no ties, so $r \neq m(a, a')$.) So, using a single comparison query, we can determine whether an agent's position is to the left or to the right of a particular midpoint. This allows us to do a binary search on the midpoints. (As an aside, if we know that each agent's position coincides with the position of one of the alternatives, then we can do a binary search on the *alternatives* to find the agent's position, using FindPeakGivenPositions. But because this is not the case in general, we need to do a binary search over the midpoints instead.) In the end, we will know the two adjacent midpoints between which the agent's position lies, which is sufficient to determine its preferences. The following example illustrates this.

**Example 3** *Suppose that the alternatives' cardinal positions (which are known to the algorithm) are as follows: $r(a) = .46, r(b) = .92, r(c) = .42, r(d) = .78, r(e) = .02$. Also suppose that the agent whose preferences we are eliciting has cardinal position $r = .52$ (which is not known to the algorithm), so that its preferences are $a \succ c \succ d \succ b \succ e$. The midpoints are, in order: $m(c, e) = .22, m(a, e) = .24, m(d, e) = .40, m(a, c) = .44, m(b, e) = .47, m(c, d) = .60, m(a, d) = .62, m(b, c) = .67, m(a, b) = .69, m(b, d) = .85$. Since the 5th midpoint (out of 10) is $m(b, e) = .47$, the algorithm first queries the agent whether it prefers b to e. The agent prefers b, so the algorithm can conclude $r > .47$ (since $r(b) > r(e)$). The 7th midpoint is $m(a, d) = .62$, so the algorithm next queries the agent whether it prefers a to d. The agent prefers a, so the algorithm can conclude $r < .62$. Finally, the 6th midpoint is $m(c, d) = .60$, so the algorithm next queries the agent whether it prefers c to d. The agent prefers c, so the algorithm can conclude $r < .60$. Now the algorithm knows that $.47 < r < .60$, and since there are no midpoints in this range, this is sufficient to deduce that the agent's preferences are $a \succ c \succ d \succ b \succ e$.*

The precise algorithm follows.

```
FindRankingGivenCardinalPositions(r)

S ← ∅
for all a, a' such that r(a) < r(a')
    S ← S ∪ {(a, a', (r(a) + r(a'))/2)}
sort S by the 3rd entry to obtain S = {(a_i, a'_i, m_i)}_{1≤i≤(m choose 2)}
l ← 0
u ← (m choose 2) + 1
while l + 1 < u {
    h ← ⌊(l + u)/2⌋
    if Query(a_h, a'_h)
        u ← h
    else
        l ← h
}
r ← (m_l + m_u)/2
return alternatives sorted by |r(a) − r|
```

**Theorem 3** FindRankingGivenCardinalPositions *correctly determines an agent's preferences, using at most $2\lceil \log(m) \rceil$ comparison queries.*

**Proof**: Correctness follows from the arguments given above. Let us consider the expression $u - l - 1$, the number of midpoints strictly between the $l$th and the $u$th. This expression starts out at $\binom{m}{2}$, and is (at least) halved by each query. Once it reaches 0, no more queries will be asked. Therefore, at most $\lceil \log(\binom{m}{2}) \rceil + 1$ queries are required. But $\log(\binom{m}{2}) = \log(m(m-1)/2) = \log(m(m-1)) - 1 < \log(m^2) - 1 = 2\log(m) - 1$. Hence FindRankingGivenCardinalPositions requires at most $2\lceil \log(m) \rceil$ comparison queries. □

This algorithm does require us to store and manage all $\binom{m}{2}$ midpoints, and because of this it does not scale computationally to extremely large numbers of alternatives, unlike the previous algorithms. It is important to remember here that elicitation cost (as measured by the number of comparison queries) is a different type of cost from computational cost. Typically, the constraints on elicitation are much more severe than those on computation. If the agent is a human, then the number of queries that she is willing to answer is likely to be very low. Even if the agent is a software agent, to answer a query, it may have to solve a hard computational problem. There are also other reasons why agents (human or software) may wish to keep the number of queries that they answer to a minimum: for example, if they are concerned about privacy, they may worry that answering more queries leaves them more exposed to a malevolent party intercepting their answers. Therefore, it is typically much more important for an elicitation algorithm to ask few queries than it is for it to be computationally very efficient. Nevertheless, in the rare settings where queries are very easily answered (presumably by a software agent), other (privacy) issues do not come into play, and the number of alternatives is very large, the computational constraint may be binding. In this case, the earlier FindRankingGivenPositions may be the preferred algorithm.

There is also an $\Omega(\log m)$ lower bound: for example, even if the agents' positions coincide with the alternatives' positions, and even without any restrictions on the type of communication (queries), an agent needs to communicate $\Omega(\log m)$ bits to identify which of the $m$ alternatives is her most preferred alternative.

## 4.2 Eliciting without Knowledge of Alternatives' Cardinal Positions

We now consider the case in which preferences are cardinally single-peaked, but the alternatives' cardinal positions are not known (at the beginning). Just as in the case of ordinally single-peaked preferences, if we do not know the alternatives' positions, then the first agent's preferences can be arbitrary and will hence require $\Omega(m \log m)$ queries to elicit. On the other hand, if we already know one agent's preferences, then we can use the algorithm for ordinally single-peaked preferences to elicit the next agent's preferences using only $O(m)$ queries, because cardinally single-peaked preferences are a special case of ordinally single-peaked preferences. However, since we are now considering cardinally single-peaked preferences, we may hope to do better than an $O(m)$ bound: after all, we saw that if the cardinal positions are known, then only $O(\log m)$ queries are required per agent. So, we might hope that after eliciting some number of agents' preferences, we will have learned enough about the cardinal positions of the alternatives that we can elicit the next agent's preferences using only $O(\log m)$ queries, or at the very least using a sublinear number of queries. Unfortunately, it turns out that this is not possible: the following result gives an $\Omega(nm)$ lower bound on the number of queries necessary to elicit $n$ agents' preferences. (The result is phrased so that it also applies to ordinally single-peaked preferences, in that case even if the (ordinal) positions are known.)

**Theorem 4** *Suppose there are n agents and m alternatives (where m is even), and the agents' preferences are known to be cardinally single-peaked, but the alternatives' positions are not known. Then, to elicit all the agents' preferences exactly, in the worst case, at least nm/2 comparison queries are necessary.*

**Remarks:** This remains true even if the alternatives' *ordinal* positions are known at the beginning—more specifically, even if each alternative's cardinal position is known at the beginning to lie in a certain interval, and these intervals do not overlap. It also remains true if (in addition) agents do not need to be queried in order—that is, it is possible to ask agent 1 some queries first, then agent 2, and then return to agent 1, *etc.*

**Proof:** Suppose that alternative $a_i$ (with $i \in \{1, \ldots, m\}$) is known (from the beginning) to lie in the interval $[k_i - 1, k_i + 1]$, where $k_i = 10 \cdot (-1)^i \lfloor (i+1)/2 \rfloor$. That is, $k_1 = -10, k_2 = 10, k_3 = -20, k_4 = 20, k_5 = -30$, *etc.* Moreover, suppose that each agent's position is known (from the beginning) to lie in the interval $[-1, 1]$. It is then clear that each agent $j$'s preferences will have the form $\{a_1, a_2\} \succ_j \{a_3, a_4\} \succ_j \{a_5, a_6\} \succ_j \ldots \succ_j \{a_{m-1}, a_m\}$ (where the set notation indicates that it is not clear which of the two is preferred). Certainly, it is *sufficient* to ask each agent, for every $i \in \{1, 3, 5, \ldots, m-1\}$, whether $a_i$ is preferred to $a_{i+1}$. The total number of queries would then be $nm/2$ (and we can ask them in any order). What we need to show, however, is that in the worst case, all of these queries are also *necessary*. To see why, suppose that the answer to every one of these queries that we ask is $a_i \succ_j a_{i+1}$ (the odd-numbered alternative is preferred). (This is certainly possible: for example, it may be that every alternative $a_i$ is at position $k_i$, and every agent is at $-1$.) Then, to be absolutely sure about every agent's complete preferences, we must ask every one of these queries, for the following reason. Suppose we have asked all queries but one: for some $j$ and odd $i$, we have not yet asked agent $j$ whether $j$ prefers $a_i$ or $a_{i+1}$. We must show that it is still possible that $a_{i+1} \succ_j a_i$. To see how this is possible, suppose that every agent $j'$ with $j' \neq j$ is at position $-1$; $j$ is at position $-0.1$; every alternative $a_{i'}$ for $i' \neq i$ is at position $k_{i'}$; and $a_i$ is at position $k_i - 1$. It is easy to see that with these positions, for any $i' \in \{1, 3, 5, \ldots, m-1\} \setminus \{i\}$, for any agent $j'$ (including $j$), we have $a_{i'} \succ_{j'} a_{i'+1}$. Moreover, even for $i$, for any agent $j' \neq j$, we have $a_i \succ_{j'} a_{i+1}$: this is because $|(k_i - 1) - (-1)| = 10(i+1)/2$, and $|k_{i+1} - (-1)| = 10(i+1)/2 + 1$. This means that these positions are consistent with all the answers to queries so far. However, with these positions, $a_{i+1} \succ_j a_i$: this is because $|(k_i - 1) - (-0.1)| = 10(i+1)/2 + .9$, and $|k_{i+1} - (-0.1)| = 10(i+1)/2 + 0.1$. Hence, we must ask the last query. □

## 5. Determining the Aggregate Ranking Only

So far, our goal has been to determine each agent's complete preferences. While it is desirable to have all this preference information, it is not always necessary. For example, our goal may simply be to determine the aggregate ranking of the alternatives. (We recall from the introduction that for single-peaked preferences, there will be no Condorcet cycles, so that the natural aggregate ranking is determined by the outcomes of the pairwise elections—assuming the number of agents is odd.) For this purpose, it is certainly *sufficient* to elicit each agent's preferences completely, but it is not clear that it is necessary. So, while Theorem 4 gives us an $\Omega(nm)$ lower bound for eliciting all agents' preferences completely in the unknown cardinal positions case (and also in the known ordinal positions case), it is not immediately clear that such a lower bound also holds when we are merely trying to determine the aggregate ranking. Unfortunately, it turns out that there is still an $\Omega(nm)$ lower bound in this case (under the same conditions as in Theorem 4). Hence, it is not possible to get away with $o(nm)$ queries—with the exception of the case of known cardinal positions, which is the only case to which this lower bound does not apply, and where we in fact only need $O(n \log m)$ queries to determine all agents' preferences completely (by using the algorithm from Subsection 4.1 for each agent).

**Theorem 5** *Suppose there are n agents and m alternatives (where n is odd and m is even), and the agents' preferences are known to be cardinally single-peaked, but the alternatives' positions are not known. Then, to determine the aggregate ranking, in the worst case, at least $(n+1)m/4$ comparison queries are necessary.*

**Remarks:** This remains true even if the alternatives' *ordinal* positions are known at the beginning—more specifically, even if each alternative's cardinal position is known at the beginning to lie in a certain interval, and these intervals do not overlap. It also remains true if (in addition) agents do not need to be queried in order—that is, it is possible to ask agent 1 some queries first, then agent 2, and then return to agent 1, *etc.*

**Proof:** The proof reuses much of the structure of the proof of Theorem 4. Again, we suppose that alternative $a_i$ (with $i \in \{1, \ldots, m\}$) is known (from the beginning) to lie in the interval $[k_i - 1, k_i + 1]$, where $k_i = 10 \cdot (-1)^i \lfloor (i+1)/2 \rfloor$. That is, $k_1 = -10, k_2 = 10, k_3 = -20, k_4 = 20, k_5 = -30$, *etc.* Moreover, again, we suppose that each agent's position is known (from the beginning) to lie in the interval $[-1, 1]$. Again, it is clear that each agent $j$'s preferences will have the form $\{a_1, a_2\} \succ_j \{a_3, a_4\} \succ_j \{a_5, a_6\} \succ_j \ldots \succ_j \{a_{m-1}, a_m\}$ (where the set notation indicates that it is not clear which of the two is preferred). More relevantly for this theorem, the aggregate ranking must also have this form. Hence, all that we need to determine is, for every $i \in \{1, 3, 5, \ldots, m-1\}$, whether more agents prefer $a_i$ to $a_{i+1}$, or vice versa. We will show that in the worst case, for each of these $m/2$ pairwise elections, we need to query at least $(n+1)/2$ agents, thereby proving the theorem. As in the proof of Theorem 4, we suppose that the answer to every one of these queries that we ask is $a_i \succ_j a_{i+1}$ (the odd-numbered alternative is preferred). Then, certainly it is sufficient to, for every $i \in \{1, 3, 5, \ldots, m-1\}$, query (any) $(n+1)/2$ agents about $a_i$ vs. $a_{i+1}$, which will result in a majority for $a_i$. But it is also necessary to ask this many queries for each of these pairwise elections, for the following reasons. Suppose that for some $i \in \{1, 3, 5, \ldots, m-1\}$, we have queried only $(n-1)/2$ agents about $a_i$ vs. $a_{i+1}$. Then, even if for all of the other pairwise elections, we have asked *all* queries, it is still possible that $a_{i+1}$ defeats $a_i$ in the pairwise election. The reason is similar to that in the proof of Theorem 4: let $J$ be the set of agents that have already been asked about $a_i$ vs. $a_{i+1}$. Suppose that every agent $j'$ with $j' \in J$ is at position $-1$; every $j \notin J$ is at position $-0.1$; every alternative $a_{i'}$ for $i' \neq i$ is at position $k_{i'}$; and $a_i$ is at position $k_i - 1$. It is easy to see that with these positions, for any $i' \in \{1, 3, 5, \ldots, m-1\} \setminus \{i\}$, for any agent $j'$ (including agents both in $J$ and outside $J$), we have $a_{i'} \succ_{j'} a_{i'+1}$. Moreover, even for $i$, for any agent $j' \in J$, we have $a_i \succ_{j'} a_{i+1}$: this is because $|(k_i - 1) - (-1)| = 10(i+1)/2$, and $|k_{i+1} - (-1)| = 10(i+1)/2 + 1$. That means that these positions are consistent with all the answers to queries so far. However, with these positions, for any $j \notin J$, $a_{i+1} \succ_j a_i$: this is because $|(k_i - 1) - (-0.1)| = 10(i + 1)/2 + .9$, and $|k_{i+1} - (-0.1)| = 10(i + 1)/2 + 0.1$. Hence, for these positions, $a_{i+1}$ defeats $a_i$ in the pairwise election. So, to rule this out, we need to ask at least one more query. □

In contrast, if we are even less ambitious and our only goal is to determine the winner (the top-ranked alternative in the aggregate ranking), then we only need to know each agent's peak: it is well known that the winner will be the median of these peaks (if the same alternative is the peak for multiple agents, then this alternative is counted multiple times in the calculation of the median). If we know at least the ordinal positions, then we can use FindPeakGivenPositions to do this using $O(\log m)$ queries per agent.

## 6. Robustness to Slight Deviations from Single-Peakedness

So far, we have assumed that the agents' preferences are always (at least ordinally, and in some cases cardinally) single-peaked. In this section, we consider the possibility that the agents' preferences are close to single-peaked, but not quite. This is easy to imagine in a political election: for example, if one of the candidates happens to be a close friend of the agent, then the agent might place this candidate high in her ranking even if they are on opposite ends of the political spectrum.

Before we get into detail on this, we should keep in mind that the case where all agents' preferences are (completely) single-peaked—the case studied before this section—is still important. While in a political election, it is likely that there are some deviations from single-peakedness (due to, for example, candidates' charisma or their positions on issues unrelated to the left-right spectrum), in other settings this seems significantly less likely. For example, if the agents are voting over numerical alternatives, such as the size of a budget, it seems likely that preferences will in fact be single-peaked.

Another issue is that many of the nice properties of single-peaked preferences will not hold for "almost" single-peaked preferences. For example, we may get Condorcet cycles again, so it is no longer clear how preferences should be aggregated. In this context, any rule for aggregating preferences is also likely to be manipulable (by declaring false preferences).

If we know the alternatives' positions, then one simple solution is to simply *require* each agent to submit single-peaked preferences (thereby forcing the agents without single-peaked preferences to manipulate, by definition). An analogous approach is often taken in combinatorial auctions, by restricting the set of valuation functions that the agents can report, for the purpose of making the winner determination problem or the elicitation problem easier. The reasoning is that the result is likely still better than not running a combinatorial auction at all. One may wonder if requiring reported preferences to be single-peaked could have some desirable effects. For example, one might argue that this forces voters in a political election to ignore the candidates' charisma (which one may argue could be desirable). A precise analysis of this is beyond the scope of this paper.

Still, it is interesting and potentially important to be able to elicit "almost" single-peaked preferences, so in this section, we study the extent to which this can be done. We can interpret the idea of the agents' preferences being "close to single-peaked" in multiple ways.

1. It may be the case that there are only a few agents that do not have single-peaked preferences (*e.g.*, few agents know any of the candidates personally, and the ones that do not will rank the candidates according to their positions on the political spectrum).

2. It may be the case that for each individual agent, that agent's preferences are close to single-peaked (*e.g.*, while an agent may know one or two of the candidates personally, there are many other candidates, and the agent will rank those remaining candidates according to their positions on the political spectrum).

In this section, we will consider mostly the first interpretation (an example illustrating why the second interpretation seems more difficult to deal with is given below). Let us suppose that a fraction $\alpha$ of the agents have preferences that are not single-peaked. One straightforward strategy is the following. Let us suppose, for now, that we know the alternatives' positions. For each agent $j$, elicit the preferences of the agent *as if* they are single-peaked. As a result, we obtain a ranking $a_1 \succ a_2 \succ \ldots \succ a_m$ of the alternatives, which we know must be $j$'s ranking *if* $j$'s preferences are in fact single-peaked. We then verify whether these are indeed $j$'s preferences, by asking $j$ to compare $a_1$ and $a_2$, $a_2$ and $a_3$, ..., and $a_{m-1}$ and $a_m$. If all of these $m-1$ queries result in the expected answer ($a_i \succ_j a_{i+1}$), then we know agent $j$'s preferences. Otherwise, we start again from the beginning, and elicit $j$'s preferences without any single-peakedness assumption, using a standard $O(m \log m)$ sorting algorithm. (Of course, in practice, we do not actually have to start completely from the beginning—we can still use the answers to the queries that we have already asked.) This leads to the following propositions.

**Proposition 1** *If a fraction $\alpha$ of the agents have preferences that are not (ordinally or cardinally) single-peaked, and the (ordinal or cardinal) positions of the alternatives are known, then we can elicit an agent's preferences using (on average) $O(m + \alpha m \log m)$ queries.*

**Proof**: FindRankingGivenPositions requires $O(m)$ queries; so does the verification step. $\alpha$ of the time the verification step fails, and we need another $O(m \log m)$ queries in that case. $\square$

We note that with this approach, there is no longer a significant advantage to having cardinally single-peaked preferences: whereas $O(\log m)$ queries are enough if we are *sure* that preferences are cardinally single-peaked (and we know the alternatives' positions), if we are not sure, we need the $m-1$ queries for the verification step.

It may be tempting to think that if an agent's preferences are almost single-peaked (in the sense of the second interpretation above), then most of the verification queries will turn out the way that we expect, so that we already know most of the agent's preferences. To see that this is not the case, suppose that the alternatives' positions are (from left to right) $a_1 < a_2 < \ldots < a_8$. Now consider a very left-wing voter who is nevertheless friends with $a_5$, resulting in the very-close-to-single-peaked preferences $a_1 \succ a_2 \succ a_3 \succ a_5 \succ a_4 \succ a_6 \succ a_7 \succ a_8$. FindRankingGivenPositions will start by running FindPeakGivenPositions, whose first query will result in the answer $a_5 \succ a_4$. As a result, FindPeakGivenPositions will conclude that the agent is a right-wing voter, and will eventually conclude that $a_5$ is the peak. FindRankingGivenPositions will then return that the preferences are $a_5 \succ a_4 \succ a_3 \succ$ $a_2 \succ a_1 \succ a_6 \succ a_7 \succ a_8$. Then, roughly half of the verification queries will not turn out the way we expect. (This example is easily generalized to any number of alternatives.)

So, let us return to the first interpretation, but now let us consider the case where the alternatives' positions are not known. For this case, we have the algorithm FindRanking-GivenOtherVote, which uses $O(m)$ queries but requires us to know another agent's preferences, which we will have to elicit first. Now, if we consider the case where not all agents' preferences are single-peaked, there are two ways in which we can get in trouble with the algorithm FindRankingGivenOtherVote: the current agent's preferences may not be single-peaked, or the earlier agent's preferences that we are using in the algorithm may not be single-peaked. In either case, the algorithm may fail to identify the current agent's preferences correctly. Again, we can address this by asking the verification queries for the current agent, and if necessary reverting to a standard sorting algorithm.

Assuming independence, the probability that both the current and the earlier agent's preferences are single-peaked is $(1-\alpha)^2$. If we are eliciting many agents' preferences, we have to be a little bit careful: if we always use the same agent's preferences as the known preferences, we run the risk that that agent's preferences are not actually single-peaked, and hence the verification step might fail for (almost) every agent. A more conservative approach is to always use the previous agent's preferences. If we assume that whether one agent's preferences are single-peaked is independent of whether this was the case for the previous agent, then indeed, $(1-\alpha)^2$ of the time both agents' preferences will be single-peaked. This leads to the following proposition.

**Proposition 2** *If a fraction $\alpha$ of the agents have preferences that are not (ordinally or cardinally) single-peaked (and this is independent from one agent to the next), and the (ordinal or cardinal) positions of the alternatives are unknown, then we can elicit an agent's preferences using (on average) $O(m + (2\alpha - \alpha^2)m \log m)$ queries (with the exception of the first agent).*

**Proof:** FindRankingGivenOtherVote requires $O(m)$ queries; so does the verification step. The verification step can only fail if either the previous or the current agent's preferences are not single-peaked, and the probability of this is $1 - (1-\alpha)^2 = 2\alpha - \alpha^2$; we need another $O(m \log m)$ queries in that case. □

(In practice, we may not want to use a different known vote every time. Rather, we may want to use the same one for a while and see if the verification step fails too often; if so, we switch to another known vote, otherwise not.)

While Propositions 1 and 2 seem to provide a good solution for the first interpretation of almost single-peaked preferences, it is less clear what to do for the second interpretation (the example above illustrates the difficulty). Still, it would be desirable to design algorithms that efficiently elicit preferences that are almost single-peaked under the second interpretation. More generally, there is a large amount of other work in preference elicitation that assumes that the preferences lie in a particular class, and it would be interesting to see if those algorithms can be generalized to deal with preferences "close" to those classes.

## 7. Conclusions

Voting is a general method for aggregating the preferences of multiple agents. Each agent ranks all the possible alternatives, and based on this, an aggregate ranking of the alternatives (or at least a winning alternative) is produced. However, when there are many alternatives, it is impractical to simply ask agents to report their complete preferences. Rather, the agents' preferences, or at least the relevant parts thereof, need to be elicited. This is done by asking the agents a (hopefully small) number of simple queries about their preferences, such as comparison queries, which ask an agent to compare two of the alternatives. Prior work on preference elicitation in voting has focused on the case of unrestricted preferences. It has been shown that in this setting, it is sometimes necessary to ask each agent (almost) as many queries as would be required to determine an arbitrary ranking of the alternatives. In contrast, in this paper, we focused on single-peaked preferences. The agents' preferences are said to be single-peaked if there is some fixed order of the alternatives, the alternatives' *positions* (representing, for instance, which alternatives are more "left-wing" and which are more "right-wing"), such that each agent prefers alternatives that are closer to the agent's most preferred alternative to ones that are further away. We first showed that if an agent's preferences are single-peaked, and the alternatives' positions are known, then the agent's (complete) preferences can be elicited using a linear number of comparison queries. If the alternatives' positions are not known, then the first agent's preferences can be arbitrary and therefore cannot be elicited using only a linear number of queries. However, we showed that if we already know at least one other agent's preferences, then we can elicit the (next) agent's preferences using a linear number of queries (albeit a larger number of queries than the first algorithm). We also showed that using a sublinear number of queries will not suffice. We also considered the case of *cardinally* single-peaked preferences—that is, each alternative and each agent has a cardinal position in $\mathbb{R}$, and agents rank alternatives by distance to their own position. For this case, we showed that if the alternatives' cardinal positions are known, then an agent's preferences can be elicited using only a logarithmic number of queries; however, we also showed that if the cardinal positions are not known, then a sublinear number of queries does not suffice. We presented experimental results for all elicitation algorithms. We also considered the problem of only eliciting enough information to determine the aggregate ranking, and showed that even for this more modest objective, a sublinear number of queries per agent does not suffice for known ordinal or unknown cardinal positions. Finally, we discussed whether and how these techniques can be applied when preferences are *almost* single-peaked. We showed how the algorithms presented earlier in the paper can be used when most agents' preferences are (completely) single-peaked. The case where each agent's preferences are almost single-peaked seems more difficult; we gave an example illustrating why.

Future research includes studying elicitation in voting for other restricted classes of preferences. The class of single-peaked preferences (over single-dimensional domains) was a natural one to study first, due to both its practical relevance and its useful theoretical properties (no Condorcet cycles and, as a result, the ability to aggregate preferences in a strategy-proof manner). Classes that are practically relevant but do not have these nice theoretical properties are still of interest, though. For example, one may consider settings where alternatives take positions in two-dimensional rather than single-dimensional space.

It is well-known that in this generalization, Condorcet cycles can occur once again. The same is true for the "almost" single-peaked settings discussed above. Nevertheless, this does not imply that efficient elicitation algorithms do not exist for these settings. Nor does it imply that such elicitation algorithms would be useless, since it is still often necessary to vote over alternatives in such settings. However, if we use a voting rule that is not strategy-proof, then we must carefully evaluate the strategic effects of elicitation. Specifically, from the queries that agents are asked, they may be able to infer something about how other agents answered queries before them; this, in turn, may affect how they (strategically) choose to answer their own queries, since the rule is not strategy-proof. This phenomenon is studied in more detail by Conitzer and Sandholm (2002).

## Acknowledgments

I thank both the AAMAS and the JAIR reviewers for their valuable feedback (the JAIR reviewers provided especially detailed and helpful feedback). Conitzer is supported under NSF award number IIS-0812113 and an Alfred P. Sloan Research Fellowship.

## Appendix A. Experimental Results for Ordinally Single-Peaked Preferences

The following experiment compares FindRankingGivenPositions, FindRankingGivenOtherVote, and MergeSort. As discussed in Subsection 2.2, MergeSort is a standard sorting algorithm that uses only comparison queries, and can therefore be used to elicit an agent's preferences without any knowledge of the alternatives' positions or of other votes. Conversely, *any* algorithm that elicits general preferences using only comparison queries can be used to solve the sorting problem. So, effectively, if we want to compare to an algorithm that does not use the fact that preferences are single-peaked, then we cannot compare to anything other than a sorting algorithm. It is conceivable that other sorting algorithms perform slightly better than MergeSort on this problem, but they all require $\Omega(m \log m)$ comparisons.

In each run, first a random permutation of the $m$ alternatives was drawn to represent the positions of the alternatives. Then, two random votes (rankings) that were single-peaked with respect to these positions were drawn. For each vote, this was done by randomly choosing a peak, then randomly choosing the second-highest ranked alternative from the two adjacent alternatives, *etc.* Each algorithm then elicited the second vote; FindRanking-GivenPositions was given (costless) access to the positions, and FindRankingGivenOtherVote was given (costless) access to the first vote. (For each run, it was verified that each algorithm produced the correct ranking.) Figure 1 shows the results (please note the logarithmic scale on the x-axis). FindRankingGivenPositions outperforms FindRankingGivenOtherVote, which in turn clearly outperforms MergeSort.

One interesting observation is that FindRankingGivenOtherVote sometimes repeats a query that it has asked before. Thus, by simply storing the results of previous queries, the number of queries can be reduced. However, in general, keeping track of which queries have been asked imposes a significant computational burden, as there are $\binom{m}{2}$ possible com-parison queries. Hence, in the experiment, the results of previous queries were not stored. FindRankingGivenPositions and MergeSort never repeat a query.

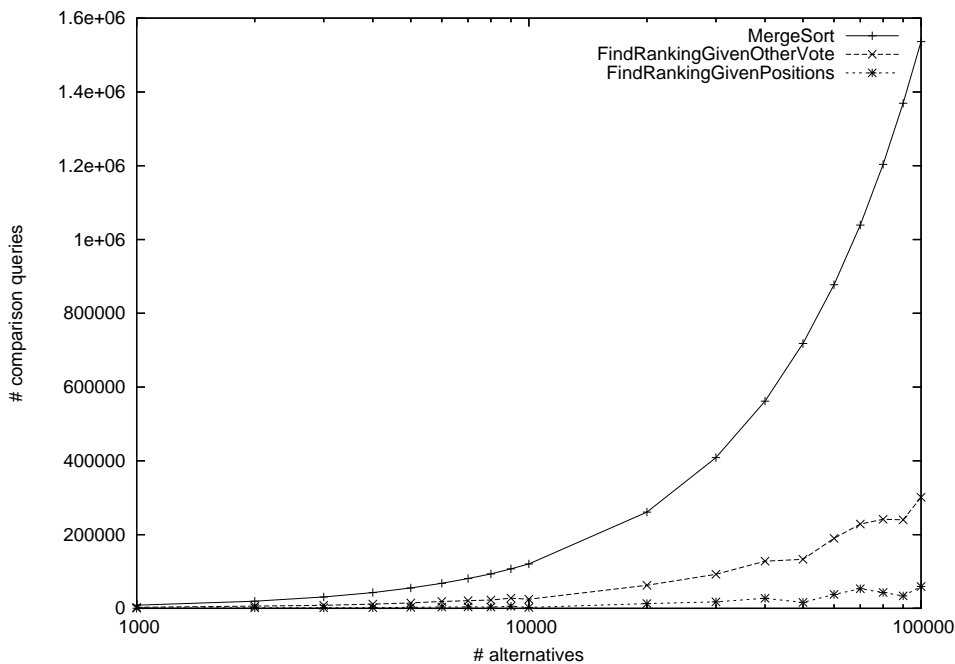

Figure 1: Experimental comparison of MergeSort, FindRankingGivenOtherVote, and Find-RankingGivenPositions. Please note the logarithmic scale on the x-axis. Each data point is averaged over 5 runs.

## Appendix B. Experimental Results for Cardinally Single-Peaked Preferences

Next, we experimentally compare FindRankingGivenCardinalPositions to FindRankingGiven-Positions. Since the former requires cardinally single-peaked preferences, we must generate preferences in a different way from Appendix A. We generate preferences in each run by drawing a cardinal position uniformly at random from $[0, 1]$ for each alternative, as well as for the agent. The agent then ranks the alternatives according to proximity to its own cardinal position. (For each run, it was verified that each algorithm produced the correct ranking.) We note that computationally, the algorithm does not scale to such large numbers of alternatives as the previous algorithms, because of the reasons mentioned earlier (managing all the midpoints). Figure 2 shows the results, clearly contrasting FindRankingGivenCardinalPositions's logarithmic nature to FindRankingGivenPositions's linear (and somewhat less predictable) nature. (Please note the logarithmic scale on the x-axis.)

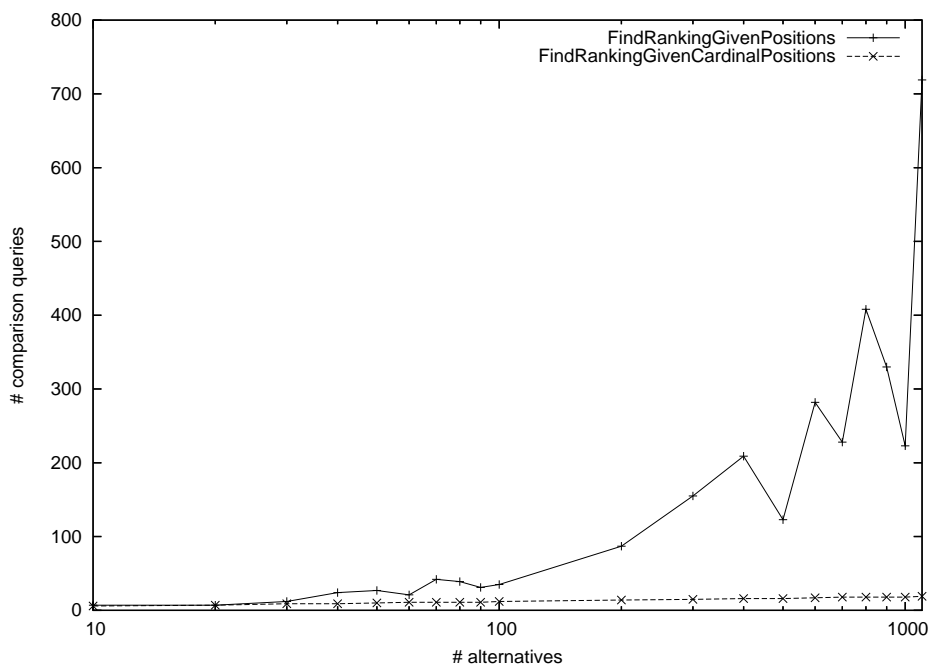

Figure 2: Experimental comparison of FindRankingGivenPositions and FindRankingGivenCardinalPositions. Please note the logarithmic scale on the x-axis. Each data point is averaged over 5 runs.